\documentclass[preprint,12pt]{elsarticle}

\usepackage{amssymb}
\usepackage{amsmath}
\usepackage{graphicx}
\usepackage{lineno}
\usepackage[colorlinks=true, allcolors=blue]{hyperref}
\usepackage{xcolor}
\usepackage{tikz}
\usetikzlibrary{arrows.meta}
\usepackage{array}
\usepackage{float}
\usepackage{tabularx}
\usepackage{ragged2e}
\journal{Physica A}

\begin{document}

\begin{frontmatter}
% Title
\title{Climate Driven Interactions Between Malaria Transmission and Diabetes Prevalence}

% Author names and affiliations
 \author[label1]{Shivank}
 \address[label1]{Department of Computer Science and Engineering, National Institute of Technology Delhi, New Delhi, 110040, India}

 \author[label1]{Anurag Singh}
 
 \author[label3]{Fakhteh Ghanbarnejad}
 \address[label3]{School of Technology and Architecture, SRH University of Applied Sciences Heidelberg, Campus Leipzig, Prager Str. 40, Leipzig, 04317, Germany }

 \author[label1]{Ajay K Sharma}

\begin{abstract}
Climate change is intensifying infectious and chronic diseases like malaria and diabetes, respectively, especially among the vulnerable populations. Global temperatures have risen by approximately $0.6^\circ$C since 1950, extending the window of transmission for mosquito-borne infections and worsening outcomes in diabetes due to metabolic stress caused by heat. People living with diabetes have already weakened immune defenses and, therefore, are at an alarmingly increased risk of contraction of malaria. However, most models rarely include both ways of interaction in changing climate conditions. In the paper, we introduce a new compartmental epidemiological model based on synthetic data fitted to disease patterns of India from 2019 to 2021. The framework captures temperature-dependent transmission parameters, seasonal variability, and different disease dynamics between diabetic and non-diabetic groups within the three-compartment system. Model calibration using Multi-Start optimization combined with Sequential Quadratic Programming allows us to find outstanding differences between populations. The odds of malaria infection in diabetic individuals were found to be 1.8–4.0 times higher, with peak infection levels in 35–36\%, as compared to 20–21\% in the non-diabetic ones. The fitted model was able to capture well the epidemiological patterns observed, while the basic reproduction number averaged around 2.3, ranging from 0.31 to 2.75 in different seasons. Given that India’s diabetic population is set to rise to about 157 million people by 2050, these findings point to a pressing need for concerted efforts toward climate-informed health strategies and monitoring systems that address both malaria and diabetes jointly.

\end{abstract}


\end{frontmatter}

% Introduce line numbers
%\linenumbers
\section{Introduction}
\begin{figure}
    \centering
    \includegraphics[width=0.7\linewidth]{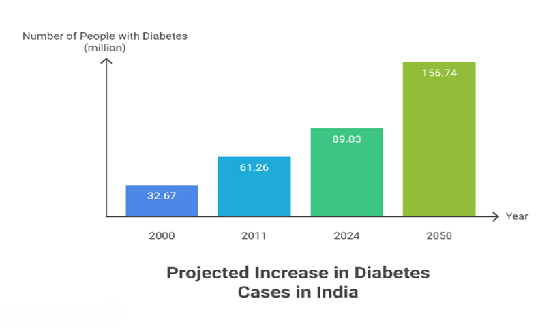}
    \caption{ Graph gives a frightening projection of diabetes in India. The population with the condition is likely to rise to a staggering 156.7 million as compared to 32.7 million in 2000, this is almost half the world, showing that a gigantic public health challenge is on the offing \cite{IDF2023}.}
    \label{fig:Diabetic_Projection}
\end{figure}

Climate change has become one of our planet's most critical challenges, and it's having a serious impact on our health. Since 1950, the world has warmed by about 0.6 $^\circ$C which has fueled a surge in diseases spread by insects, especially malaria. The parasites that cause malaria are carried by Anopheles mosquitoes, and these mosquitoes are very sensitive to temperature. As a result, warmer weather is making it easier for malaria to spread \cite{wahyudin2018sir}. At the same time, climate change is making life harder for people with diabetes. They are hit harder by heat stress, air pollution, and wilder rainfall patterns \cite{ratter2023diabetes}. Malaria is a preventable disease, but it continues to be a heavy burden for low- and middle-income countries, causing thousands of deaths and years of disability \cite{GBD2016_Causes_of_Death, GBD2016_Disease_and_Injury}. The World Health Organization (WHO) estimated there would be over 260 million malaria cases and nearly 597,000 deaths worldwide by 2023, mostly in places with few resources \cite{WHO_Malaria}. Concerning situation is the link researchers are finding between chronic health conditions and infectious diseases. A large study in Sweden (1995–2015) discovered that obesity and diabetes are major risk factors for severe malaria in adults. The study clearly showed that patients with severe malaria were much more likely to have other chronic diseases, with diabetes being a particularly strong risk factor \cite{wyss2017obesity}. Climate change has fasten the process by extreme heat, polluted air, and severe weather events can disrupt the health of someone with diabetes for months or years, making it harder to control their blood pressure and blood sugar. People with diabetes often have compromised immune systems, making them more likely to get dangerously sick from malaria. On top of that, simply living in a hotter climate is leads to more people developing diabetes in the first place \cite{ratter2023diabetes}. As temperatures climb, diabetes rates are expected to soar. As seen in Figure \ref{fig:Diabetic_Projection}, projections show India could become one of the world's most diabetic nations by 2050 \cite{IDF2023}. Climate change is helping malaria spread in regions where healthcare is already limited, precisely where growing numbers of people with diabetes are struggling to cope. Changing rainfall and hotter temperatures create perfect breeding grounds for mosquitoes, while other factors like population density and mosquito biting rates all shape how fast the disease spreads.

Understanding the complex interplay between vectors, their environments, and disease transmission has been essential for predicting and controlling the spread of vector-borne diseases. Liu-Helmersson et al. investigated the vectorial capacity of Aedes aegypti, highlighting the profound impact of temperature on global dengue epidemic potential, though their model relies heavily on theoretical calculations but lacks validation using real-world datasets \cite{liu2014vectorial}. Similarly, Ghosh et al. investigated the effects of temperature and nutritional stress on Anopheles stephensi, revealing changes in vectorial competence and life-history traits, but with limited analysis of how these temperature-induced changes influence malaria transmission in urban settings under climate change scenarios \cite{ghosh2024effects}.

Epidemiological modeling has been pivotal in understanding and forecasting vector-borne diseases. Wang and Mulone introduced a model describing disease spread between two patches, highlighting the reproduction number as a threshold indicator for disease persistence \cite{wang2007epidemic}. Nana-Kyere et al. used nonlinear equations for compartmental models to show the changing dynamics between host and malaria, tracking susceptible, vaccinated, exposed, infectious, and recovered groups for humans, and aquatic larvae, susceptible adults, and infectious adults for mosquitoes, using $R_0$ to determine disease-free and stability states in Ghana, though the model does not account for climate change and environmental factors \cite{nana2024mathematical}. Temperature, as a key environmental factor, plays a crucial role in shaping disease transmission. Arquam et al. incorporated temperature variations into the classical SIR model, demonstrating how climatic conditions can influence the dynamics of vector-borne epidemics \cite{arquam2020integrating}. Rogers emphasized the integration of spatial and environmental data to improve the accuracy and relevance of predictive models, with logistic regression and discriminant analysis widely used to model vector distributions, accounting for environmental factors such as temperature and habitat suitability \cite{rogers2006models}.

Mathematical models have proven valuable for understanding how diabetes develops and spreads through populations. The Susceptible-Diabetes-Complications (SDC) framework, for example, considers lifestyle choices, genetic factors, and social interactions to forecast diabetes rates and complications \cite{Widyaningsih2018, boutayeb2006non}. Interestingly, diabetes and climate change influence each other in both directions. Rising temperatures affect how our bodies process glucose and respond to insulin, leading to more hospital admissions among people with diabetes \cite{boutayeb2006non}. A study in China's Shandong Province found clear links between temperature, humidity, and diabetes-related deaths among elderly patients \cite{Zheng2024}. Ratter-Rieck and colleagues showed that heat waves particularly harm diabetic individuals, increasing their hospitalization rates. Their research suggests that just a 1°C rise in temperature could lead to 100,000 additional diabetes cases each year in the United States alone \cite{ratter2023diabetes}.

The relationship between diabetes and infectious diseases, such as malaria, has also garnered attention. People with diabetes have weaker immune systems, making them more susceptible to infections. A systematic review indicated a positive association between type 2 diabetes and increased risk of malaria, including its severity \cite{CarrilloLarco2019, Danquah2010}. Studies in Ghana revealed that diabetic individuals had a 46\% greater risk of malaria infection, though there remains a gap in understanding the long-term effects of compromised immune systems on malaria transmission and the distinction between diabetic and non-diabetic individuals in terms of malaria risk \cite{Danquah2010}. The inclusion of climatic factors in disease modeling has advanced understanding of vector-borne diseases, with SIR models incorporating temperature and humidity utilized to analyze dengue fever transmission, demonstrating how environmental conditions influence the basic reproduction number and disease dynamics \cite{Nur2018}.

Although studies have explored the effects of climatic variables on VBDs and diabetes independently, limited research has addressed the integration of diabetic population dynamics with vector population dynamics under environmental modulation. Diabetes research often examines glucose-insulin dynamics, such as the Bergman minimal model \cite{bergman1979quantitative}, while more recent population-level models incorporate risk factors and progression rates between pre-diabetes and diabetes states \cite{king1998global}. A deeper understanding of these interactions is essential to identify shifting disease patterns and to develop effective strategies for mitigating the dual impact of climate change and disease dynamics.

\section{Methodology for interaction between VBD dynamics and pre-existing disease. }

In the proposed methodology for investigating the complex interplay between vector-borne infectious disease dynamics and non-communicable disease progression is developed, using malaria and diabetes as case studies. The approach integrates climate influence, physiological mechanisms, and epidemiological patterns to represent the disease interactions.

\subsection{ Dataset Description }

As comprehensive, real-world time-series data for malaria-diabetes comorbidity is not available, we generated a synthetic dataset for the purpose of model calibration. This dataset was parameterized using real-world epidemiological statistics from the SCB Medical College study in Odisha, a Ghana case study, and WHO reports to ensure it realistically reflects the observed disease dynamics. The study revealed significantly higher mortality rates among diabetic patients compared to non-diabetics, with 35.18\% (19 out of 54) deaths in the diabetic group versus 13.69\% (20 out of 146) deaths in the non-diabetic group, indicating a 2.57-fold increased risk of mortality in diabetic patients. Clinical findings shows more severe malaria in diabetic patients such as longer infection duration and higher complication rates were used to set differential recovery rates and transmission susceptibility values in the synthetic dataset\cite{thatoi2018diabetes}. The malaria incidence rate of 2.1 per 1,000 population, as reported by the World Health Organization (WHO) India , was used to guide the baseline transmission intensity \cite{whoMalaria2024}. Data were simulated over a 36-month period with daily resolution to capture both seasonal variation and long-term trends in disease transmission.
Population structure was based on the ICMR INDIAB national survey, which reported an 11.4\% diabetes prevalence age $\geq 20$ years\cite{anjana2023icmr}, resulting in a modeled population of 80{,}000 diabetic and 920{,}000 non-diabetic individuals. The mosquito population was set at a 2:1 mosquito-to-human ratio, consistent with entomological studies \cite{mcclure2025relating}, giving a total of 2{,}000{,}000 vectors. Transmission probabilities were assigned as 0.65 for diabetic and 0.50 for non-diabetic individuals, reflecting the 46\% higher malaria risk in diabetic patients reported in a Ghana case study \cite{Danquah2010}. A mosquito has a 75\% chance of becoming infected when it bites an infected diabetic individual, compared to a 50\% chance when biting an infected non-diabetic individual. with a mosquito mortality rate fixed at $1/14$ per day \cite{mcclure2025relating}.

Seasonal variation in biting rate was modeled using a sinusoidal function, peaking post-monsoon. Recovery rates assumed an average infection duration of 90 days. Gaussian noise (15\% for diabetic, 20\% for non-diabetic compartments) was added to simulate observational uncertainty and natural variability. The complete dataset structure with variable descriptions, ranges, and units is detailed in Table~\ref{tab:variables}.

\begin{table}[htbp]
\centering
\caption{Description of dataset variables.}
\label{tab:variables}
\begin{tabular}{ll}
\hline
\textbf{Description} & \textbf{Range} \\
\hline
Time points ($t$) & 0 -- 1080 \\
Susceptible diabetic individuals ($S_D$) & 75{,}000 -- 80{,}000 \\
Infected diabetic individuals ($I_{MD}$) & 0 -- 5{,}000 \\
Susceptible non-diabetic individuals ($S_{-D}$) & 870{,}000 -- 920{,}000 \\
Infected non-diabetic individuals ($I_M$) & 0 -- 50{,}000 \\
Susceptible vectors  ($S_v$) & 1{,}900{,}000 -- 2{,}000{,}000 \\
Infected vectors  ($I_v$) & 0 -- 100{,}000 \\
Observed infected diabetics with noise (observed $ {I}_{MD}$) & 0 -- 5{,}000 \\
Observed infected non-diabetics with noise (observed ${I}_M$) & 0 -- 50{,}000 \\
\hline
\end{tabular}
\end{table}

\subsection{Compartmental model}
\begin{figure}[h]
    \centering
    \includegraphics[width=1\linewidth]{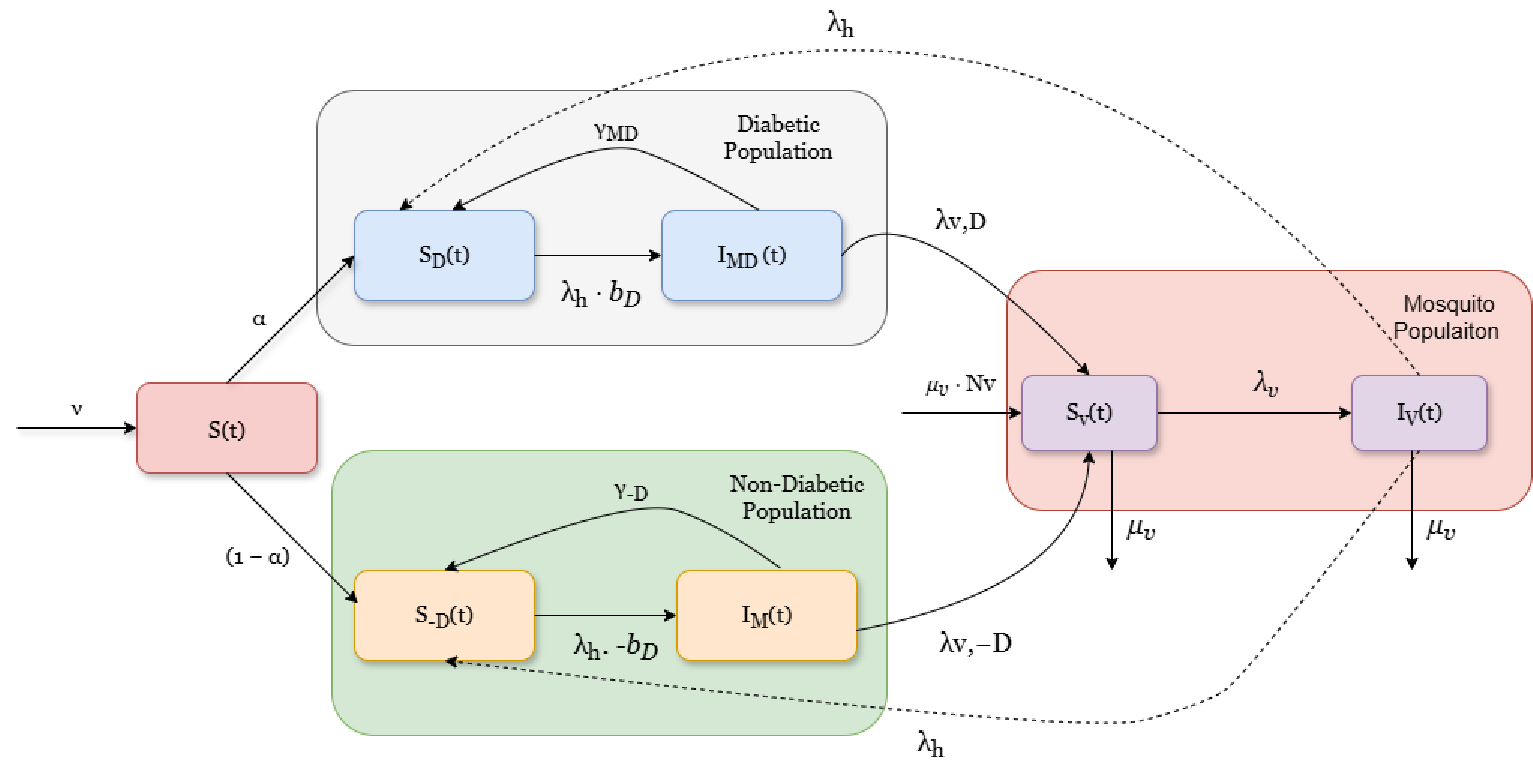}
    \caption{The diagram shows how malaria spreads differently among people with and without diabetes. A general susceptible population $S(t)$ splits into diabetic ( $\alpha$) and non-diabetic ( $1-\alpha$) subpopulations with different transmissions. Diabetic individuals have transmission rate $\lambda_h \cdot b_D$ and recovery rate $\gamma_{MD}$, while non-diabetics follow similar flows with rate $\lambda_h \cdot b_{-D}$ and recovery $\gamma_{-D}$. Mosquito populations transition from susceptible $S_v(t)$ to infected $I_v(t)$ at rate $\lambda_v$, with mortality $\mu_v$ in both compartments. Cross-transmission occurs bidirectionally: infected vectors transmit to humans via force of infection $\lambda_h$ represented by dashed arrows, While infected humans transmit back to vectors through rates $\lambda_{v,D}$ for diabetic individuals and $\lambda_{v,-D}$ for non-diabetic individuals.
 The model demonstrates how diabetes creates differential transmission and recovery dynamics within a vector-borne disease system while maintaining homogeneous mixing between populations.}

    \label{fig:compartmental diagram}
\end{figure}

To investigate the complex epidemiological interactions between malaria and diabetes, we developed a compartmental model that captures the bidirectional relationship between diabetic and malaria shown in Figure \ref{fig:compartmental diagram}.

\subsection{Malaria-Diabetes Epidemiological Model}

In the section, we have developed a mathematical model that comprises six compartmental equations that describe the dynamics of malaria transmission between diabetic and non-diabetic human populations and mosquito vectors. The model also assumes homogeneous mixing between vector and human populations, meaning all individuals have equal probability of contact regardless of location. The model assumes fixed total population sizes for both vector and human populations throughout the study period. 
The model stratifies the human population by diabetes status, 
where $\alpha = \tfrac{N_D}{N^{\text{total}}_h} = 0.08$ represents the proportion 
of diabetic individuals and $(1-\alpha) = 0.92$ represents non-diabetics population (-D). Vector 
population balance is maintained through a recruitment rate 
$\nu = \mu_v \cdot N_v$, offsetting vector mortality to keep the total vector 
population constant at $N_v = 2{,}000{,}000$. The human population remains stable 
with no births or deaths, with individuals transitioning only between susceptible 
and infected states within their respective subpopulations.
For the diabetic population, the relationship $
S_D + I_{MD} = N_D = 80{,}000$ ensures that the total number of diabetic people remains constant, with individuals only moving between susceptible and infected states. Similarly, for the non-diabetic population, $
S_{-D} + I_{M} = N_{-D} = 920{,}000 $ maintains population conservation. The total vector population is also conserved with $
S_v + I_v = N_v = 2{,}000{,}000 
$ individuals, where mosquitoes transition between susceptible and infected compartments.
The model maintains conservation of population by assuming that individuals only move between compartments from susceptible to infected or vice versa without being added or removed from the system through births, deaths, or migration.
To reflect biological realism, the human population is considered stable during the study period, with no demographic changes, while the vector population undergoes continuous births and deaths that are perfectly balanced to maintain a constant total mosquito population.

\begin{equation}
\frac{dS_D}{dt} = -\lambda_h \cdot b_D \cdot S_D + \gamma_{MD} \cdot I_{MD}
\label{eq:sus_diab}
\end{equation}
where $\lambda_h$ is the force of infection from vectors to humans, $b_D = 0.65$ is the transmission probability for diabetics, $S_D$ represents the susceptible diabetic population, $\gamma_{MD} = 1/120$ days$^{-1}$ is the recovery rate for diabetics, and $I_{MD}$ is the infected diabetic population.

\begin{equation}
\frac{dI_{MD}}{dt} = \lambda_h \cdot b_D \cdot S_D - \gamma_{MD} \cdot I_{MD}
\label{eq:Inf_diab}
\end{equation}

\begin{equation}
\frac{dS_{-D}}{dt} = -\lambda_h \cdot b_{-D} \cdot S_{-D} + \gamma_{-D} \cdot I_{M}
\label{eq:Non_diab_sus}
\end{equation}
 $b_{-D} = 0.50$ is the transmission probability for non-diabetics, $S_{-D}$ represents the susceptible non-diabetic population, and $\gamma_{-D} = 1/60$ days$^{-1}$ is the recovery rate for non-diabetics.

\begin{equation}
\frac{dI_{M}}{dt} = \lambda_h \cdot b_{-D} \cdot S_{-D} - \gamma_{-D} \cdot I_{M}
\label{eq:Non_diab_Inf}
\end{equation}

\begin{equation}
\frac{dS_v}{dt} = \mu_v \cdot N_v - \lambda_{v,D} \cdot S_v - \lambda_{v,-D} \cdot S_v - \mu_v \cdot S_v
\label{eq:sus_vector}
\end{equation}
where $\mu_v = 1/14$ days$^{-1}$ is the vector mortality rate, $N_v = 2{,}000{,}000$ is the total vector population, $\lambda_{v,D}$ is the force of infection from diabetic humans, $\lambda_{v,-D}$ is the force of infection from non-diabetic humans, and $S_v$ is the susceptible vector population.

\begin{equation}
\frac{dI_v}{dt} = (\lambda_{v,D} + \lambda_{v,-D}) \cdot S_v - \mu_v \cdot I_v
\label{eq:Inf_vector}
\end{equation}
where $I_v$ is the infected vector population.
The forces of infection are defined as below,

\begin{equation}
\lambda_h = \frac{a(t) \cdot I_v}{N_v} \tag{7}
\end{equation}
where $\lambda_h$ represents the rate at which susceptible humans become infected 
through contact with infected vectors, proportional to the biting rate $a(t)$ 
and the fraction of infected vectors $I_v/N_v$.

\begin{equation}
\lambda_{v,D} = \frac{a(t) \cdot c_D \cdot I_{MD}}{N^{\text{total}}_h} \tag{8}
\end{equation}
where $\lambda_{v,D}$ represents the rate at which susceptible vectors become 
infected from diabetic humans, determined by the biting rate $a(t)$, transmission 
probability from diabetic to vector $c_D$, and the proportion of infected diabetics 
$I_{MD}/N^{\text{total}}_h$ in the human population.

\begin{equation}
\lambda_{v,-D} = \frac{a(t) \cdot c_{-D} \cdot I_{M}}{N^{\text{total}}_h} \tag{9}
\end{equation}
where $\lambda_{v,-D}$ represents the rate at which susceptible vectors become 
infected from non-diabetic humans, determined by the biting rate $a(t)$, transmission 
probability from non-diabetic to vector $c_{-D}$, and the proportion of infected 
non-diabetics $I_{M}/N^{\text{total}}_h$ in the human population.

\begin{table}[ht]
\centering
\small
\caption{Model Parameters and Descriptions}
\label{tab:model_parameters}

\begin{tabularx}{\textwidth}{|l| >{\RaggedRight}X |l|l|}
\hline
\textbf{Parameters} & \textbf{Description} & \textbf{Value} & \textbf{Units} \\
\hline
$N_D$ & Total diabetic population size & $80{,}000$ & individuals \\
$N_{-D}$ & Total non-diabetic population size & $920{,}000$ & individuals \\
$N_v$ & Total vector population size & $2{,}000{,}000$ & individuals \\
$N_h^{\text{total}}$ & Total human population size & $1{,}000{,}000$ & individuals \\
$b_D$ & Transmission probability from vector to diabetic & $0.65$ & dimensionless \\
$b_{-D}$ & Transmission probability from vector to non-diabetic & $0.50$ & dimensionless \\
$c_D$ & Transmission probability from diabetic to vector & $0.75$ & dimensionless \\
$c_{-D}$ & Transmission probability from non-diabetic to vector & $0.50$ & dimensionless \\
$\gamma_{MD}$ & Recovery rate for diabetic individuals & $1/120 \approx 0.0083$ & days$^{-1}$ \\
$\gamma_{-D}$ & Recovery rate for non-diabetic individuals & $1/60 \approx 0.0167$ & days$^{-1}$ \\
$\mu_v$ & Vector mortality rate & $1/14 \approx 0.0714$ & days$^{-1}$ \\
$a_{\text{mean}}$ & Mean biting rate & $0.1$ & days$^{-1}$ \\
$a_{\text{amp}}$ & Seasonal amplitude of biting rate & $ 0.8 $ & dimensionless \\
\hline
\end{tabularx}
\end{table}

\subsubsection{ Basic Reproduction Number }

The basic reproduction number, $R_0$, represents the expected number of secondary infections produced by a single infected individual in a completely susceptible population. For vector-borne diseases with multiple host populations, $R_0$ is derived using the next-generation matrix approach, %which systematically separates new infections from other state transitions to capture the transmission dynamics between different population groups.%

The infected compartments in our malaria-diabetes system are $I_{MD}$, $I_{M}$ , and $I_v$, leading to the formation of two key matrices: $F$  and $V$. The matrix $F$ describes the rate at which new infections occur between compartments, where diabetic individuals become infected by vectors at rate $\frac{a \cdot b_D \cdot S_D^0}{N_v}$, non-diabetic individuals are infected by vectors at rate $\frac{a \cdot b_{-D} \cdot S_{-D}^0}{N_v}$, and vectors acquire infections from diabetic and non-diabetic humans at rates $\frac{a \cdot c_D \cdot S_v^0}{N_{h,\text{total}}}$ and $\frac{a \cdot c_{-D} \cdot S_v^0}{N_{h,\text{total}}}$, respectively.

At the disease-free equilibrium (DFE), all individuals are in susceptible states, giving us $S_D^0 = N_D = 80{,}000$, $S_{-D}^0 = N_{-D} = 920{,}000$, and $S_v^0 = N_v = 2{,}000{,}000$, where $N_{h,\text{total}} = N_D + N_{ND} = 1{,}000{,}000$. Using these equilibrium values, the matrices are constructed as:

\begin{equation}
\mathbf{F} =
\begin{pmatrix} 
0 & 0 & \frac{a \cdot b_D \cdot N_D}{N_v} \\
0 & 0 & \frac{a \cdot b_{-D} \cdot N_{-D}}{N_v} \\
\frac{a \cdot c_D \cdot N_v}{N_{h,\text{total}}} & \frac{a \cdot c_{-D} \cdot N_v}{N_{h,\text{total}}} & 0
\end{pmatrix}
\end{equation}

\begin{equation}
\mathbf{V} =
\begin{pmatrix} 
\gamma_{MD} & 0 & 0 \\
0 & \gamma_{-D} & 0 \\
0 & 0 & \mu_v
\end{pmatrix}
\end{equation}

The basic reproduction number is mathematically defined as the spectral radius (largest eigenvalue) of the matrix $\mathbf{F}\mathbf{V}^{-1}$. For vector-borne diseases with multiple host types, $R_0$ can be elegantly expressed as the geometric mean of the reproduction numbers for each direction of transmission:

\begin{equation}
R_0 = \sqrt{R_{0,HV} \cdot R_{0,VH}}
\end{equation}

Here, $R_{0,HV}$ represents the expected number of vectors infected by one infected human throughout their infectious period, and $R_{0,VH}$ represents the expected number of humans infected by one infected vector throughout its lifespan. These directional reproduction numbers are calculated using population-weighted averages of the transmission parameters to account for heterogeneity in the human population:

\begin{equation}
R_{0,HV} = \frac{a \cdot c_{\text{eff}} \cdot N_v}{\mu_v \cdot N_{h,\text{total}}}, \quad
R_{0,VH} = \frac{a \cdot b_{\text{eff}} \cdot N_{h,\text{total}}}{\gamma_{\text{eff}} \cdot N_v}
\end{equation}

The effective transmission parameters are calculated as:

\begin{align}
b_{\text{eff}} &= \frac{N_D \cdot b_D + N_{-D} \cdot b_{-D}}{N_{h,\text{total}}} = \frac{52{,}000 + 460{,}000}{1{,}000{,}000} = 0.512 \\
c_{\text{eff}} &= \frac{N_D \cdot c_D + N_{-D} \cdot c_{-D}}{N_{h,\text{total}}} = \frac{60{,}000 + 460{,}000}{1{,}000{,}000} = 0.52 \\
\gamma_{\text{eff}} &= \frac{N_D \cdot \gamma_D + N_{-D} \cdot \gamma_{-D}}{N_{h,\text{total}}} = \frac{664 + 15{,}364}{1{,}000{,}000} = 0.016
\end{align}

Substituting into the geometric mean expression:

\begin{equation}
R_0 = a \sqrt{\frac{b_{\text{eff}} \cdot c_{\text{eff}}}{\mu_v \cdot \gamma_{\text{eff}}}} = a \sqrt{\frac{0.26624}{0.001143}} = a \cdot 15.26
\end{equation}

Using the parameter value $a = 0.1$ days$^{-1}$ from our model:

\begin{equation}
R_0 = 0.1 \cdot 15.26 = 1.526
\end{equation}

This value indicates that the disease will persist and spread in the population since $R_0 > 1$, meaning each infected individual will generate approximately 1.526 secondary infections on average in a completely susceptible population.

\subsubsection{Seasonal Variation in \texorpdfstring{$R_0$}{R₀}}

% The seasonal variation in the biting rate parameter is modeled as:

% \[
% a(t) = a_{\text{mean}} \left(1 + a_{\text{amp}} \cos\left(\frac{2\pi(t/30.4 - 10)}{12}\right)\right)
% \]

% with $a_{\text{mean}} = 0.1$ and $a_{\text{amp}} = 0.8$. This causes $R_0$ to fluctuate throughout the year:

% \begin{itemize}
%     \item Minimum $R_0$: $0.1 \cdot (1 - 0.8) \cdot 15.26 = 0.305$
%     \item Maximum $R_0$: $0.1 \cdot (1 + 0.8) \cdot 15.26 = 2.747$
% \end{itemize}

% These fluctuations generate the temporal dynamics characteristic of seasonal malaria transmission patterns.

The seasonal variation in the biting rate parameter is modeled as

\[
a(t) = a_{\text{mean}} \left(1 + a_{\text{amp}} \cos\left(\frac{2\pi(t/30.4 - 10)}{12}\right)\right)
\]

with $a_{\text{mean}} = 0.1$ and $a_{\text{amp}} = 0.8$, where each parameter reflects important aspects of malaria transmission ecology. The mean biting rate of $0.1$ per day represents the average frequency of mosquito-human contact under normal environmental conditions, corresponding to approximately $3$ bites per month per mosquito-human pair. This estimate is based on field studies showing that 
Anopheles mosquitoes typically require blood meals every $2$--$3$ days for reproduction, though not all feeding attempts result in successful human contact. The seasonal amplitude of $0.8$ results in a variation of $80\%$ around the mean, which reflects the dramatic changes in mosquito populations commonly found within tropical areas. In these habitats, the density of vectors may vary by many orders of magnitude (within the range of dry and rainy seasons) because of variations in the availability of breeding sites, development rates that are dependent on temperature, and the survival of adult mosquitoes. The cosine function brings in periodic oscillations of a smooth nature, which reflect natural seasonal cycles. Daily units can be converted into months by the time conversion factor of $t/30.4$, as malaria seasonality is most clearly seen on a monthly basis. A negative offset of $-10$ is used to align transmission peaks with the months that occur after peak rainfall, as an indication of the biological reality that mosquito populations typically peak two to three months after the beginning of the rainy season because of the larval growth period. The periodicity of 12 months per year guarantees repetition of the cycle after every 12 months, as is observed globally in terms of malaria transmission seasons.

\section{Results and Discussion}

In the following section, the tools and methodology to be applied to the analysis are discussed. In addition, the dataset analysis is achieved through combining the prevalence of malaria and diabetes, which combines epidemiological data with environmental data to understand the disease patterns and the impact of climate. Lastly, the findings on the simulations of the proposed model will be given and these will prove to be effective in examining the interactions among these variables.

\subsection{Experimental setup}
The model simulations were coded in Python with the use of \texttt{scipy.}\\\texttt{integrate.solve\_ivp} to solve the mathematical equations. The population in the study was a representation of both diabetics and non-diabetics, with mosquito vectors maintained at realistic proportions relative to the human population. Several years of simulations with daily time steps were performed, enabling us to capture changes in transmission patterns across seasons. To determine the most appropriate parameters for biting rates, seasonal variation, recovery rates, and initial conditions, we employed \texttt{scipy.optimize}\\\texttt{.minimize} with the L-BFGS-B method, ensuring that all values remained within biologically realistic bounds. Gaussian noise was introduced to represent real-world measurement errors, with higher noise levels for non-diabetic cases, as these are more likely to be misdiagnosed. Model validation was conducted throughout the process to ensure that the results were biologically meaningful and that total population numbers were conserved. The obtained results were then visualized using \texttt{matplotlib}.

\subsection{Result}

\begin{figure}[H]
    \centering
    \includegraphics[width=1\linewidth]{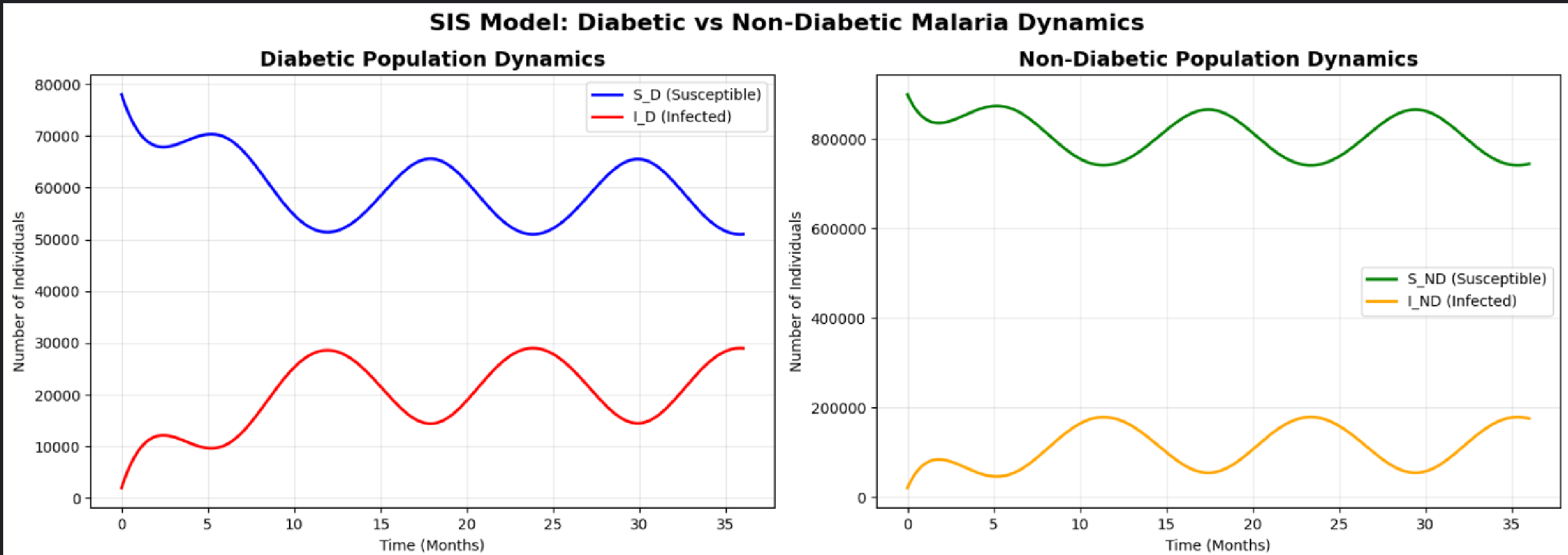}
    \caption{The figure shows the difference in the spread of malaria between diabetic and non-diabetic populations in the 36-month period with a Susceptible-Infected-Susceptible (SIS) model. The trends of the two groups are clearly different. Strong seasonal swings are seen among the diabetic population of 80,000 people. The weakened group is reduced to an average of between 50,000 to 52,000 during the peak transmission as the population of infected individuals increases to between 28,000 to 29,000, implying that at these times there is a 35--36\% infection rate among diabetics. Compared to this, the non-diabetic population of 920,000 people exhibits synchronized seasonal variations with less dramatic changes in comparison. The vulnerable population is between 750,000 and 850,000 people, with the infected cases ranging from 180,000 to 190,000, which represents lower infection rates of 20--21\%. The model has three full cycles of epidemics per year. The relative infection peaks, oscillations, and epidemic volatility are higher in the diabetic population compared to the non-diabetics. These variations are a factor of higher transmission probability ($0.65$ in diabetics versus $0.50$ in non-diabetics) and slower recovery, with an average of 120 days of infection compared with 60 days of infection in non-diabetics. Even with these distinctions, both populations exhibit synchronized seasonal patterns, which demonstrate that both populations are influenced by common environmental factors.
.}
    \label{fig:Compatmental_plot}
\end{figure}

The mechanistic model successfully captured distinct epidemiological dynamics between 
diabetic and non-diabetic populations over the 36 month study period shown in Figure \ref{fig:Compatmental_plot}. Both populations exhibited synchronized seasonal oscillations driven by shared 
environmental drivers, with the diabetic population of 80,000 individuals 
showing pronounced fluctuations. $S_{D}$ declined from 78,000 to 50,000--52,000 during peak transmission, while $I_{MD}$ reached 
28,000--29,000 individuals (35--36\% prevalence). The non-diabetic population showed 
less dramatic relative changes, with $I_{M}$ peaking at 
180,000--190,000 (20--21\% prevalence). The three complete annual cycles demonstrate 
higher epidemic volatility in diabetics due to combined effects of higher 
transmission probability ($b_D=0.65$ vs $b_{-D}=0.50$) and slower recovery rates.

\begin{figure}[H]
    \centering
    \includegraphics[width=1\linewidth]{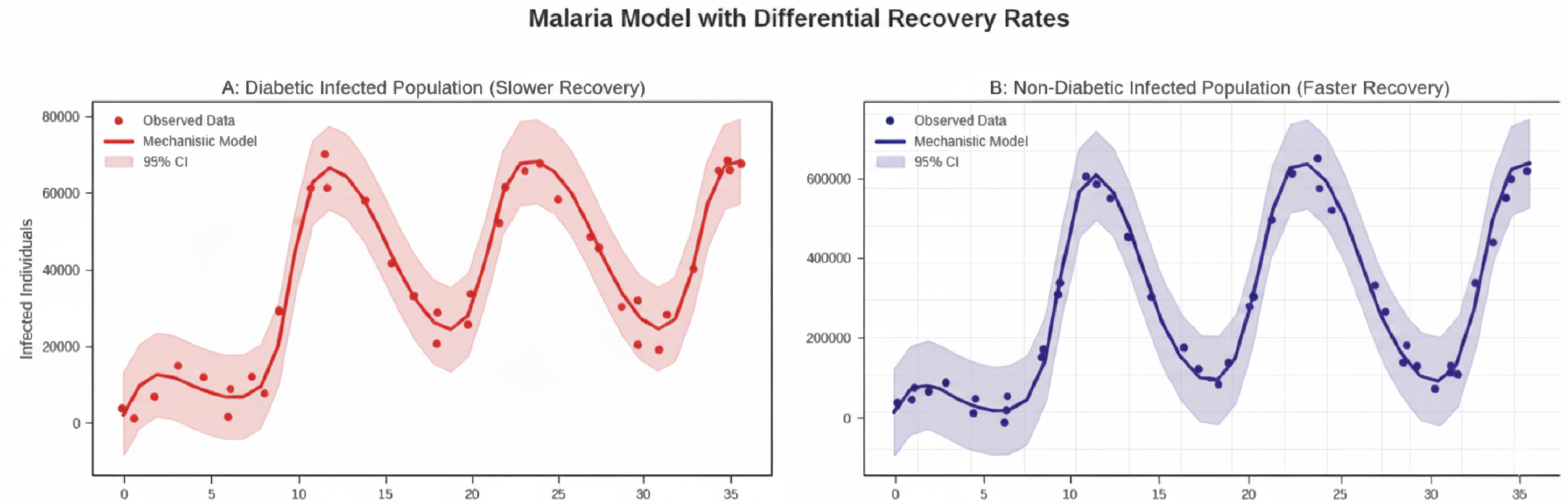}
    \caption{The figure shows how   the model adjusted using different recovery rates for each group. The left panel tracks infected diabetics ($I_{MD}$) who recover more slowly ($\gamma_{MD} = \tfrac{1}{120}$ per day), while the right panel shows infected non-diabetics ($I_{M}$) recovering faster ($\gamma_{-D} = \tfrac{1}{60}$ per day). The red and blue dots are our synthetic data with added Gaussian noise, based on real clinical statistics. Solid lines show how well the model fits this data, with shaded regions marking 95\% confidence intervals. The tight fit confirms the model successfully captures seasonal patterns, elevated baseline infections, and sustained epidemic dynamics that align with clinical observations. Data points plotted are month wise.}

    \label{fig:Model_Dynamics}
\end{figure}

The model also captures the seasonal transmission  with differential recovery rates patterns observed in both diabetic and non-diabetic populations over the 36-month study period as shown in Figure~\ref{fig:Model_Dynamics}. The model uses mean biting rate of 0.099, seasonal amplitude of 0.810, while setting recovery rates based on clinical evidence. Diabetic individuals demonstrated a slower recovery rate of 0.0082 per day, corresponding to an infection duration of approximately 122.2 days, while non-diabetic individuals exhibited a faster recovery rate of 0.0163 per day, equivalent to 61.2 days. When we compared model predictions against the observed synthetic data, we found strong alignment. Importantly, diabetic populations maintained consistently higher baseline infection levels and experienced more prolonged epidemics, which can be attributed directly to their extended infection periods.

\begin{figure}[H]
    \centering
    \includegraphics[width=0.8\textwidth]{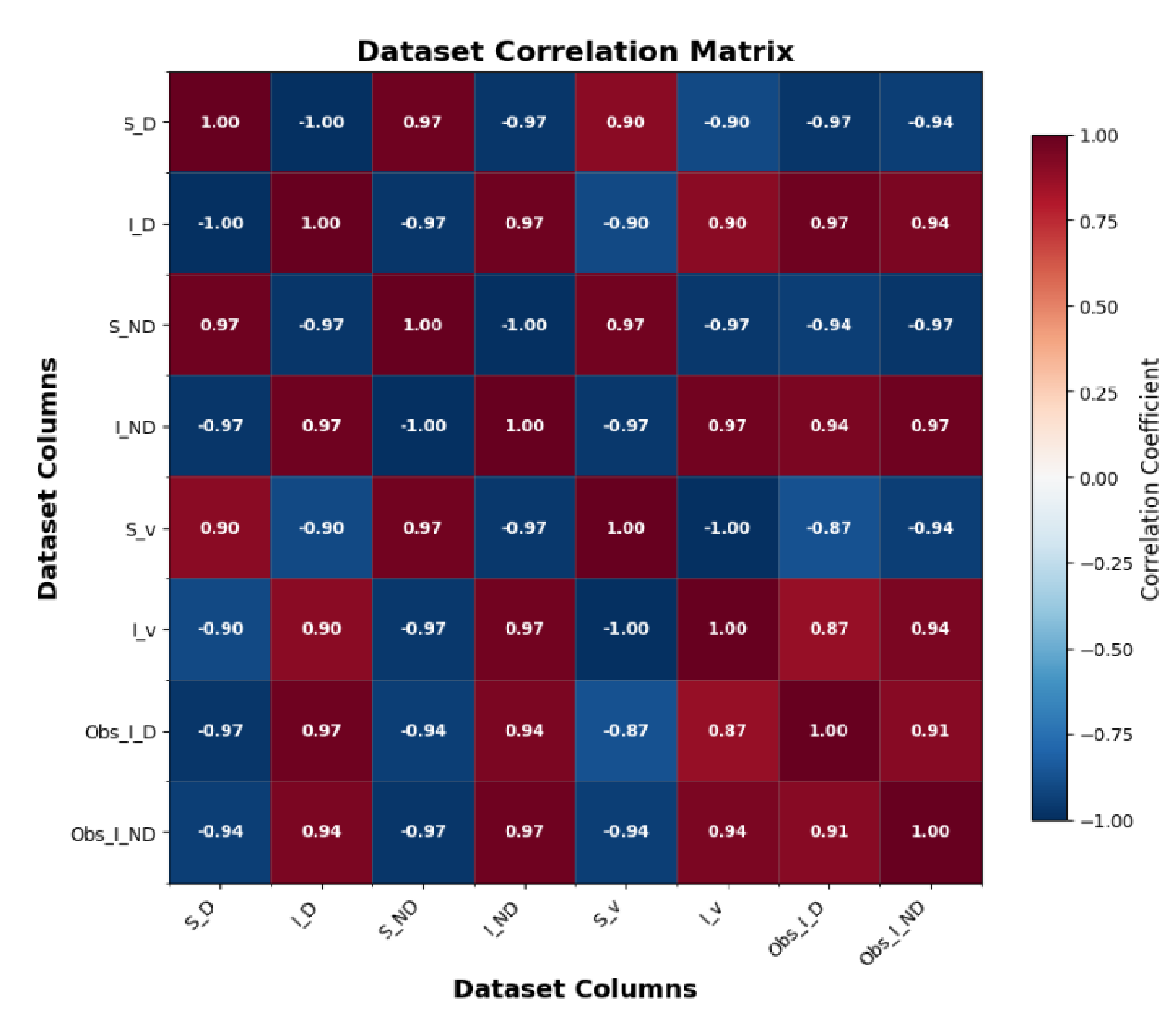} 
    \caption{The heatmap displays the relationships between different disease compartments across diabetic and non-diabetic individuals and mosquito populations over a 3-year study period. The model parameters set included the mean biting rate of $0.099$, seasonal amplitude of $0.810$, and recovery rates between groups: for diabetics at $0.0082$ per day (122.2-day infection duration) and for non-diabetics at $0.0163$ per day (61.2-day duration). Several key patterns arise within the correlation structure. Within each population group, susceptible and infected compartments display perfect negative correlations ($-1.00$), which intuitively makes sense given the closed population assumption of SIS models where individuals simply move back and forth between these two states. Strong positive correlations ($0.90$–$0.97$) across all infected populations imply that epidemics among diabetic and non-diabetic groups rise and fall in tandem. Both populations respond to the same seasonal transmission drivers, explaining the synchronized pattern, even though non-diabetics clear infections roughly twice as fast as diabetics. Simulated predictions from the model show high agreement with observed data ($0.94$–$0.97$), confirming that real underlying relationships are well captured even in the presence of measurement noise. The color gradient ranges from dark blue ($-1.00$), through white (no correlation), to dark red ($+1.00$).}
    \label{fig:heatmap}
\end{figure}

Strong correlations were observed among all model components, demonstrating the interconnected nature of malaria transmission dynamics between populations as shown in Figure~\ref{fig:heatmap}. Perfect negative correlations ($-1.00$) between susceptible and infected compartments within each population reflected the closed population constraint inherent to SIS models. Also, strong positive correlations ($0.90$--$0.97$) were found among all infected populations, indicating synchronized epidemic dynamics driven by shared seasonal transmission patterns, despite the two-fold faster recovery rate in non-diabetics compared to diabetics. The high correlation between simulated and observed data ($0.94$--$0.97$) confirmed robust model performance and validated the mechanistic relationships despite observational noise.

%figure 5

\begin{figure}[h]
    \centering
    \includegraphics[width=0.8\linewidth]{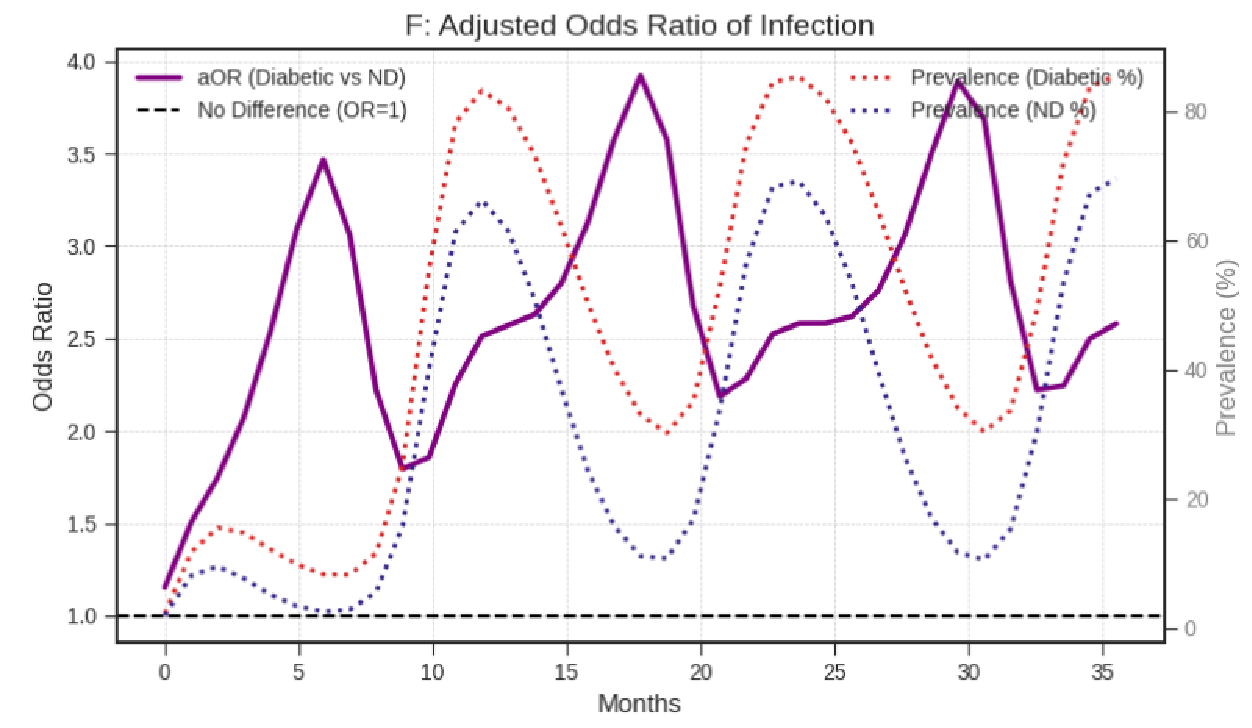}
    \caption{
As indicated in the figure, diabetic patients were still susceptible to malaria infection after 36 months, and there were clear cyclic patterns of malaria infection after 12 months in 3 seasonal cycles. The solid purple line is the adjusted odds ratio (aOR) that always remains higher than the reference line (OR = 1, black dashed) and varies between 1.8 and 4.0 with peaks close to 4.0 when the periods of high transmission occur. As indicated in figure that diabetic people are almost four times increased to be infected as compared to non-diabetic individuals in these peak seasons. These findings are supported by the prevalence data (red dotted line is diabetics and blue dotted line is non-diabetics) where the prevalence of infection of diabetics increases to 80--85\% during high seasonal seasons in contrast to 60--70\% in non-diabetics, and the prevalence of infection of non-diabetics is 10--15\% compared to 5--10\%, respectively. The fact that the peaks of vulnerability due to diabetes are synchronized indicates that diabetes is a significant comorbidity of malaria that requires targeted public health interventions, especially in the seasons of high risk of malaria.
}
    \label{fig:aoR}
\end{figure}

The aOR analysis revealed that diabetic individuals consistently faced elevated malaria infection risk throughout the entire study period (Figure~\ref{fig:aoR}). The aOR fluctuated between $1.8$ and $4.0$, never falling to the reference line ($\text{OR} = 1$), indicating persistent elevated risk in the diabetic population. Risk disparities intensified during high transmission seasons, with peak aOR values approaching $4.0$, meaning diabetic individuals had nearly four times the odds of infection compared to non-diabetic individuals. Prevalence data supported these findings, showing that diabetic individuals exhibited infection rates of $80$--$85\%$ during peak seasons compared to $60$--$65\%$ in non-diabetic individuals. During low transmission periods, prevalence remained elevated in diabetics ($15$--$20\%$) compared to non-diabetics ($10$--$15\%$). The data demonstrated clear 12-month cyclical patterns repeated three times, with the greatest risk disparities occurring during periods of heightened transmission intensity. These results confirm that diabetes represents a significant comorbidity that amplifies malaria vulnerability across multiple seasonal cycles.

\section{Conclusions and future directions}

We developed an integrated epidemiological framework to model the bidirectional interactions between malaria transmission and diabetes prevalence under climate forcing. Using a three-compartment mathematical model calibrated with synthetic data reflecting Indian disease patterns from 2019 to 2021, we quantified substantially elevated malaria risk among diabetic individuals. Our findings show that individuals with diabetes face 1.8--4.0 times higher odds of malaria infection compared to non-diabetics, and the length of time they remain infected is about 122 days versus just 61 days in those without diabetes. The longer duration of infectiousness creates lingering disease reservoirs that continue to sustain transmission despite overall prevalence decline, and peak infection rates rise to 35--36\% amongst diabetics compared to 20--21\% in non-diabetics. Seasonal fluctuations in the basic reproduction number allow identification of critical climate-indexed intervention windows when disease transmission wanes during unfavorable conditions but can spread explosively when environmental conditions become optimal.

We have demonstrated in our work that climate change, diseases transmitted by the vectors, and non-communicable diseases are mutually related risks and solutions cannot be provided in isolation. Since the rise in temperatures across the globe and the rise in diabetes, the communities are becoming more and more exposed to convergent epidemics with diseases reinforcing each other. These two issues require combined surveillance measures to monitor cross-cutting disease burdens, seasonal interventions that are specific to most vulnerable groups to protect them during peak epidemic times and climate-adaptive approaches to the population-based health that should be developed long before the dual epidemic pressures overwhelm health systems in resource-constrained situations.

Despite the fact that the existing model provides fresh understanding of the relationship between malaria and diabetes during climatic stress, it can be expanded in several ways to increase its predictive ability and policy impact. Subsequent improvements can focus on real-life hospital data relating glycemic control to malaria phenotypes, age-stratified compartments to account vulnerability over life stages, explicit mosquito cycle development at varying temperatures, stochastic components that more realistically reflect the uncertainty, acquired immunity, a wider variety of climate factors including rainfall distribution and heat stress, and extension to other co-circulating diseases vectors including dengue and chikungunya.

% \section*{Acknowledgments}
% [Add your acknowledgments here]

% \section*{Author Contributions}
% [Add author contributions information here]

% \section*{Funding}
% [Add funding information here]

% \section*{Declaration of Competing Interest}

% References
\bibliographystyle{elsarticle-num}
\bibliography{ref}

\end{document}